\documentclass[pdflatex,sn-mathphys-num]{sn-jnl}

\usepackage{graphicx}%
\usepackage{multirow}%
\usepackage{amsmath,amssymb,amsfonts}%
\usepackage{amsthm}%
\usepackage{mathrsfs}%
\usepackage[title]{appendix}%
\usepackage{xcolor}%
\usepackage{textcomp}%
\usepackage{manyfoot}%
\usepackage{booktabs}%
\usepackage{algorithm}%
\usepackage{algorithmicx}%
\usepackage{algpseudocode}%
\usepackage{listings}%
\usepackage{pdflscape}
\usepackage{afterpage}
\usepackage{comment}

\newcommand{\A}{{\rm A}}
\newcommand{\B}{{\rm B}}
\newcommand{\Ab}{{\rm Ab}}
\newcommand{\Ac}{{\rm Ac}}
\newcommand{\msun}{{\rm M}_\odot}

\newcommand{\au}{{\rm AU}}
\newcommand{\tot}{{\rm tot}}
\newcommand{\age}{{\rm age}}

\theoremstyle{thmstyleone}%
\theoremstyle{thmstyletwo}%

\theoremstyle{thmstylethree}%

\begin{document}

\title{Planet formation and long-term stability in a very eccentric stellar binary}


\author[1]{\fnm{Jakob} \sur{Stegmann}}\email{jstegmann@mpa-garching.mpg.de}

\author[2,3]{\fnm{Evgeni} \sur{Grishin}}

\author[4,1,5]{\fnm{Cole} \sur{Johnston}}

\author[6,7]{\fnm{Nora L.} \sur{Eisner}}

\author[1]{\fnm{Stephen} \sur{Justham}}

\author[1,8]{\fnm{Selma E.} \sur{de Mink}}

\author[9,10]{\fnm{Hagai B.} \sur{Perets}}

\affil[1]{\orgname{Max Planck Institute for Astrophysics}, \orgaddress{\street{Karl-Schwarzschild-Str. 1}, \postcode{85748} \city{Garching}, \country{Germany}}}

\affil[2]{\orgdiv{School of Physics and Astronomy}, \orgname{Monash University}, \orgaddress{\city{Clayton}, \state{VIC} \postcode{3800}, \country{Australia}}}

\affil[3]{\orgname{OzGrav: Australian Research Council Centre of Excellence for Gravitational Wave Discovery},  \orgaddress{\city{Clayton}, \state{VIC} \postcode{3800}, \country{Australia}}}

\affil[4]{\orgdiv{Department of Physics}, \orgname{University of Surrey}, \orgaddress{\city{Guildford}, \state{Surrey}, \postcode{GU2 7XH}, \country{UK}}}

\affil[5]{\orgdiv{Institute of Astronomy}, \orgname{KU Leuven}, \orgaddress{\street{Celestijnenlaan 200D}, \postcode{3001} \city{Leuven}, \country{Belgium}}}

\affil[6]{\orgdiv{Center for Computational Astrophysics}, \orgname{Flatiron Institute}, \orgaddress{\street{162 Fifth Avenue}, \city{New York City}, \state{NY} \postcode{10010}, \country{USA}}}

\affil[7]{\orgdiv{Department of Astrophysical Sciences}, \orgname{Princeton University}, \orgaddress{\city{Princeton}, \state{NJ} \postcode{08544}, \country{USA}}}

\affil[8]{\orgname{Ludwig-Maximilians-Universität München}, \orgaddress{\street{Geschwister-Scholl-Platz 1}, \postcode{80539} \city{München}, \country{Germany}}}

\affil[9]{\orgdiv{Physics Department}, \orgname{Technion - Israel Institute of Technology}, \postcode{3200003} \orgaddress{\city{Haifa}, \country{Israel}}}

\affil[10]{\orgname{Astrophysics Research Center of the Open University (ARCO)}, \postcode{4353701} \orgaddress{\city{Raanana}, \country{Israel}}}

\abstract{Planets orbiting one of the two stars in a binary are vulnerable to gravitational perturbations from the other star. Particularly, highly eccentric companion stars risk disrupting planetary orbits, such as in the extreme system TOI\,4633 where close encounters between the companion and a gas giant planet in the habitable zone make it one of the most fragile systems discovered so far. Here, we report that TOI\,4633’s planet likely survived these encounters throughout the system’s age by orbiting retrograde relative to the binary, stabilised by the Coriolis force. Using direct $N$-body simulations, we show it otherwise tends to collide with the binary stars or becomes free-floating after getting ejected. A retrograde planetary orbit has profound implications for TOI\,4633's formation and evolution, suggesting an extraordinary history where its eccentric companion was likely randomly captured after planet formation in a single-star system. Alternatively, if stars and planet are born in situ from the same gas clump, we show the planet must have formed at sub-snow-line distances, contrary to the conventional core-accretion model. Our study highlights the importance of considering the long-term stability ($\gtrsim\rm Gyr$) of planets in eccentric binaries and demonstrates that the mere existence in such dynamically hostile environments places strong constraints on their orbital configuration and formation.}

\maketitle

\begin{figure}
    \centering
    \includegraphics[width=\linewidth]{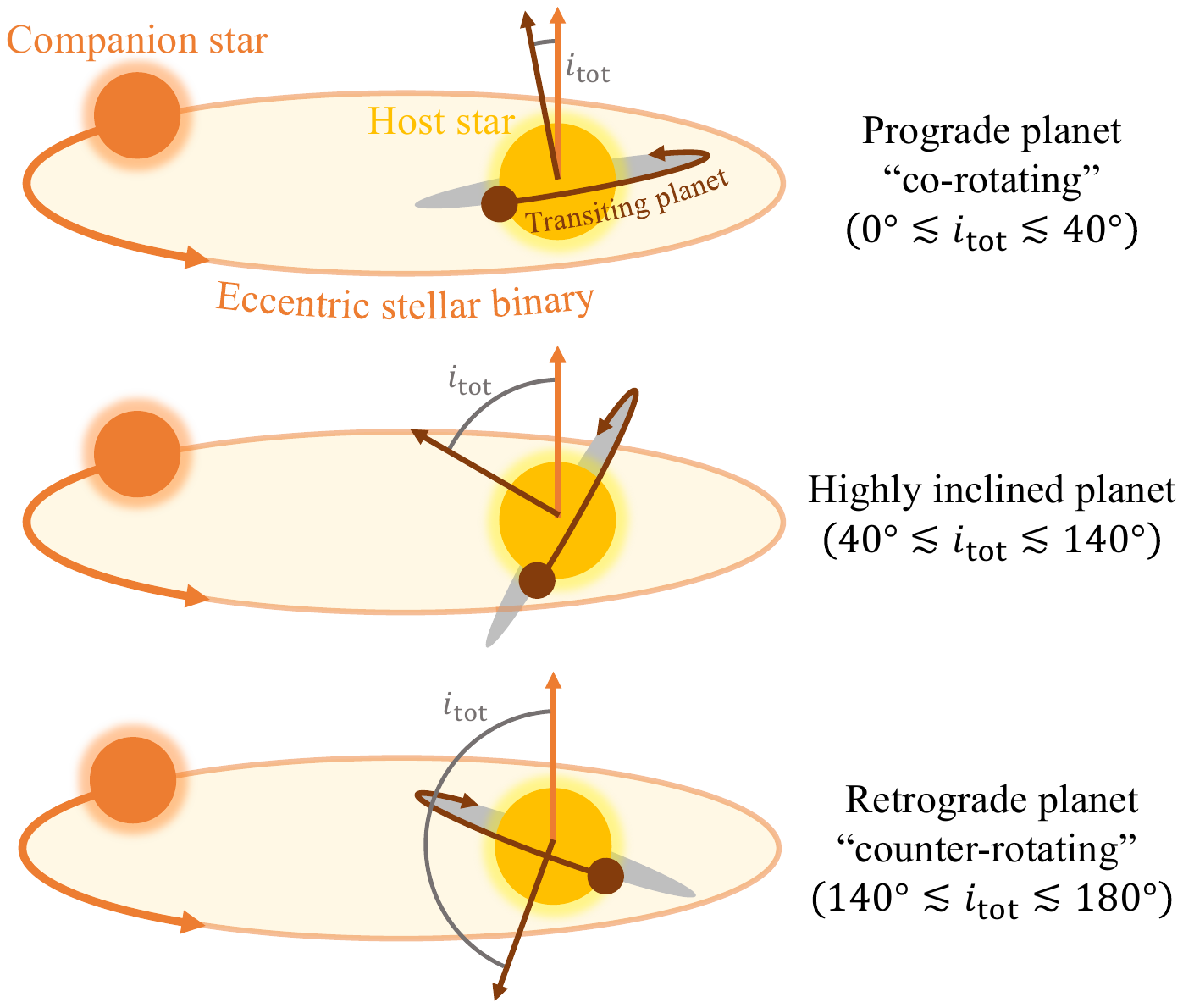}
    \caption{Schematic overview of the eccentric binary system TOI\,4633 hosting an s-type planet in the habitable zone. The planet was discovered via the transit detection method which leaves the spatial orientation of its orbit unknown. It may either be (a) prograde, i.e., the planet and the companion star share the same sense of rotation, (b) highly inclined, or (c) retrograde as illustrated in the three panels. We find the planet likely to orbit retrograde in order to survive the periodic gravitational perturbations throughout the system's lifetime of $t_\age=1.3\pm0.3\,\rm Gyr$.}
    \label{fig:sketch}
\end{figure}

TOI\,4633 is a remarkable system that was recently discovered as part of the Citizen Science Project Planet Hunters TESS \citep{Eisner2024}. It consists of two solar-like stars moving around one another on a highly eccentric orbit ($e_\B=0.91\pm0.03$) and a transiting mini-Neptune (TOI\,4633c) in the habitable zone around one of the stars (TOI\,4633A). The large eccentricity causes the stellar binary companion (TOI\,4633B) to closely encounter the planet at a periapsis $r_{p,\B}\approx4.4\,\au$ (see sketch in Figure~\ref{fig:sketch}), once every binary period of about $230\,\rm yr$. Of more than two hundred planets discovered in stellar binaries, these close encounters make TOI\,4633c one of the most dynamically fragile planets known to date.

\begin{figure}
    \centering
    \includegraphics[width=\linewidth]{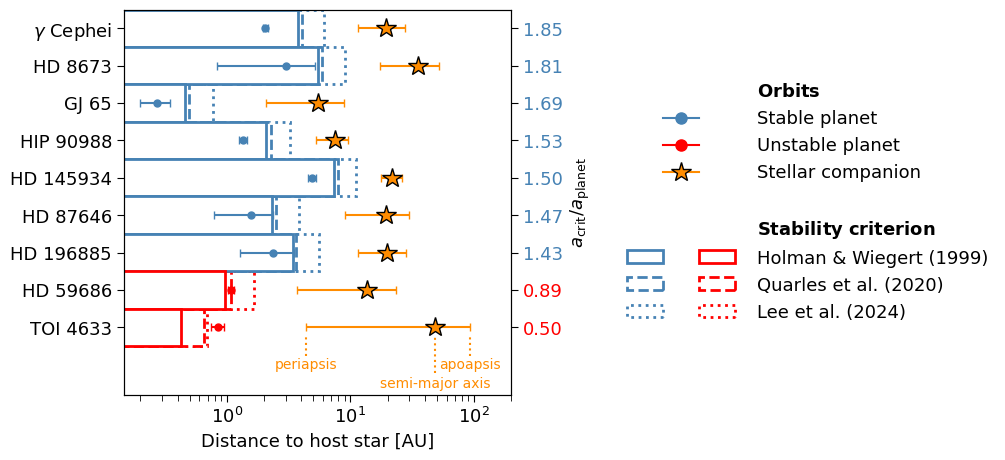}
    \caption{Orbital hierarchy of previously discovered stellar binaries with s-type planets that are close to being dynamically unstable. We show all systems with $a_{\rm crit}/a_{\rm planet}<2$ with the semi-major axis of the (outermost) planet $a_{\rm planet}$ \citep{Holman1999} contained in The Extrasolar Planets Encyclopaedia \cite{Encyclopaedia} and sort them according to their presumed stability by $a_{\rm crit}/a_{\rm planet}$. Orange star symbols indicate the semi-major axis $a_\B$ of the stellar binary while horizontal lines extend from its periapsis $r_{p,\B}=a_\B(1-e_\B)$ to apoapsis $r_{a,\B}=a_\B(1+e_\B)$. Similarly, blue and red circles and lines indicate the semi-major axis, periapsis, and apoapsis of the planet, respectively. Blue and red bars extend to the critical semi-major axis $a_{\rm crit}$ for the planetary orbit above which it would not be dynamically stable according to \citet{Holman1999} (solid), \citet{Quarles2020} (dashed), and \citet{Lee2024} (dotted). For the criterion of \citet{Holman1999}, we show if a planet is stable (blue) or not (red) and indicate the numerical value on the right.} 
    \label{fig:s-type-Sample}
\end{figure}

It is confounding how such a planet could have formed and survived throughout TOI\,4633's age of $t_{\age}=1.3\pm0.3\,{\rm Gyr}$ \citep{Eisner2024}. The periodic gravitational perturbation from the closely encountering stellar companion should pose a significant risk to the stability of TOI\,4633c’s orbit, potentially resulting in its ejection from the system or a collision with the host star. Indeed, previous stability criteria which were inferred from a grid of direct three-body integrations of single s-type\footnote{S-type (circumstellar) planets such as TOI\,4633c orbit either one of the two stars of a binary (see Fig.~\ref{fig:sketch}), whereas p-type (circumbinary) planets orbit the binary as a whole at a much larger distance.} planets in binary systems  suggest that TOI\,4633c is on the verge of destruction. Figure \ref{fig:s-type-Sample} shows that TOI\,4633c's semi-major axis $a_\Ac=0.847\pm0.061\,\au$ exceeds the widely used critical threshold $a_{\rm crit}$ of dynamical stability from \citet{Holman1999}. While their grid-based interpolation for stability is formally inapplicable to TOI\,4633c since they only consider co-planar orbits with eccentricities up to $e_\B\le0.8$, \citet{Eisner2024} conducted simulations to specifically recover dynamically stable configurations of TOI\,4633. As a result, the planet could only be stable if the mutual inclination angle $i_\tot$ between its orbit around TOI\,4633A and the orbital plane of the stellar binary TOI\,4633AB is between $0^\circ\lesssim i_\tot\lesssim40^\circ$ (prograde) or $140^\circ\lesssim i_\tot\lesssim180^\circ$ (retrograde), whereas highly inclined orbits ($40^\circ\lesssim i_\tot\lesssim140^\circ$) become quickly unstable (see Figure~\ref{fig:sketch} for a schematic overview). The fragility of TOI\,4633 is even more concerning given the common expectation that planets of that mass ($m_{\rm c}\approx50\,\rm M_\oplus$, see Supplementary Table~\ref{tab:orbital-parameters} for a list of all parameters) need to form beyond the snow line at $a_\Ac\gtrsim3.0\,\au$ \cite{hayashi1981,Ida2005} where the temperature is low enough for volatile compounds like water to condense into solid ice grains. Beyond the snow line the detrimental effect of the encountering companion would be even stronger, or it may truncate the protostellar disc around the host star preventing the planet to form.

Previous pioneering studies to assess the stability of general or specific s-type planets largely rely on direct $N$-body simulations that only cover several to tens of Myr \cite{Holman1999,Pilat-Lohinger2002,David2003,Pilat-Lohinger2003,Barnes2004,Musielak2005,Fatuzzo2006,Marzari2007,Quarles2016,Trifonov2018,Quarles2020,Quarles2024,Eisner2024,Lee2024}. This is only a small fraction $\lesssim1\,\%$ of the estimated age $t_{\age}\gtrsim\rm Gyr$ of the stars in TOI\,4633 and other typical low-mass stellar hosts of observed planets. Simulating systems like TOI\,4633, which are on the verge of destruction, up to a maximum integration time $t_{\max}\ll t_{\age}$ could yield false positive stable solutions. The perturbation from the stellar companion could accumulate to destabilise the planet between $t_{\max}$ and $t_{\age}$. Here, we present a suite of direct $N$-body simulations \cite{rebound} of TOI\,4633 which were conducted for a maximum integration time $t_{\max}=t_{\age}$ to evolve different realisations of TOI\,4633 (see Methods) which are consistent with the observed parameters (Supplementary Table~\ref{tab:orbital-parameters}). As a default, TOI\,4633 is arranged as a 2+1 triple composed of the two stars TOI\,4633A and B, as well as the planet TOI\,4633c orbiting star A (Figure~\ref{fig:sketch}) and explore plausible variants below. 
The planet TOI\,4633c was discovered via the transit detection method, i.e., by causing dips in stellar brightness when it passes in front of its host stars and blocks some of its light. As illustrated in Figure~\ref{fig:sketch}, the dip in stellar brightness does not reveal at which angle of incidence the planet moves in front of the star, which effectively leaves the mutual inclination $i_\tot$ between the planet orbit and the binary plane the only fundamentally unconstrained parameter of TOI\,4633 (see Methods). Therefore, we initialise the simulations of TOI\,4633 by choosing different values of $i_\tot$ between $0^\circ$ and $180^\circ$ (see Methods).
All simulations are terminated if either the planet survives by reaching the maximum integration time $t_{\max}=t_{\age}$ or if the planet becomes unstable due to a collision with a star or by getting ejected from the system.

\begin{figure}
    \centering
    \includegraphics[width=0.97\linewidth]{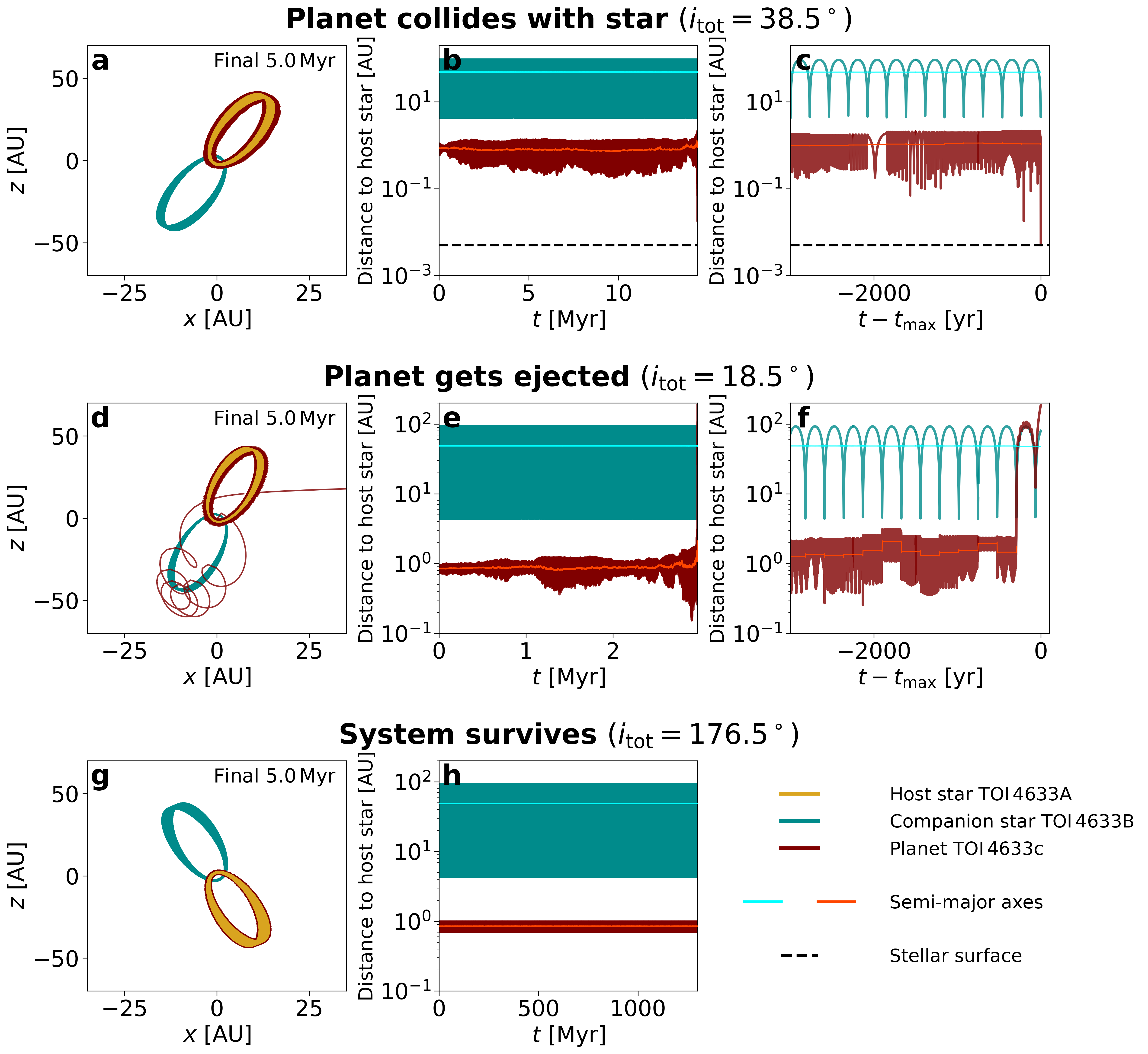}
    \caption{Three examples for the evolution of TOI\,4633 in a 2+1 configuration. The system in the upper row (panels a~--~c) starts from a mutual inclination $i_\tot=38.5^\circ$ leading to a collision of planet TOI\,4633c with its host star TOI\,4633A after $t_{\max}\approx1.4\times10^7\,\rm yr$. The system in the middle row (d~--~f) starts from a mutual inclination $i_\tot=18.5^\circ$ resulting in the ejection of the planet after $t_{\max}\approx3.0\times10^6\,\rm yr$. Only in the lower row (g and~h) does the system remain intact throughout its estimated age $t_{\max}=t_{\age}=1.3\times10^9\,\rm yr$ after starting from $i_\tot=176.5^\circ$. The leftmost panels (a, d, and~g) display the projected trajectories in the centre-of-mass frame of the two stars (purple and red) and the planet (cyan) during the final $5.0\times10^6\,\rm yr$. The middle panels (b, e, and~h) show the relative separations of the companion star TOI\,4633B and the planet with respect to the host star as a function of time. Light colours indicate the semi-major axes of their orbits around the host and the dashed line corresponds to the combined radius $r_\A+r_\Ac$ of the host star and the planet. For the non-surviving systems, the rightmost panels (c and~f) contain a zoom-in of the last $3\times10^3\,\rm yr$ before the planetary collision and ejection, respectively.}
    \label{fig:examples}
\end{figure}
Figure~\ref{fig:examples} presents example systems that illustrate the three qualitatively different evolutionary outcomes by showing the trajectories of the stars and the planet and their relative separations as a function of time. In the first two examples (panels~a~--~c and d~--~f) the planets start out on prograde orbits ($i_\tot<40^\circ$). In these cases, the periodic forcing from the stellar companion at periapsis impulsively induces a chaotic evolution of the planets around their host which eventually leads to a collision and ejection, respectively. Notably, the planet in the second example (panels~d~--~f) does get ejected and becomes a free-floating planet \cite{Miret-Roig2022} after it had been captured by the stellar companion and followed its motion for one orbit. We find this ``star hopping" \cite{Kratter2012} to be a common event preceding a planetary ejection. The planet in the third example (panels~g~and~h) is initialised on a retrograde orbit ($i_\tot>140^\circ$). It remains largely unaffected by the stellar companion and survives the entire evolution for $t_{\max}=t_{\age}$. Meanwhile, in all three examples the stellar binary remains largely unaffected by the presence of the planet except that it causes a notable precession of the binary periapsis by a few degrees  (left panels~a, d, and g), which is consistent with a precession timescale of $T_{\rm prec, B} \approx 170\,\rm Myr$ (see Methods). 

\begin{figure}
    \centering
    \includegraphics[width=0.97\linewidth]{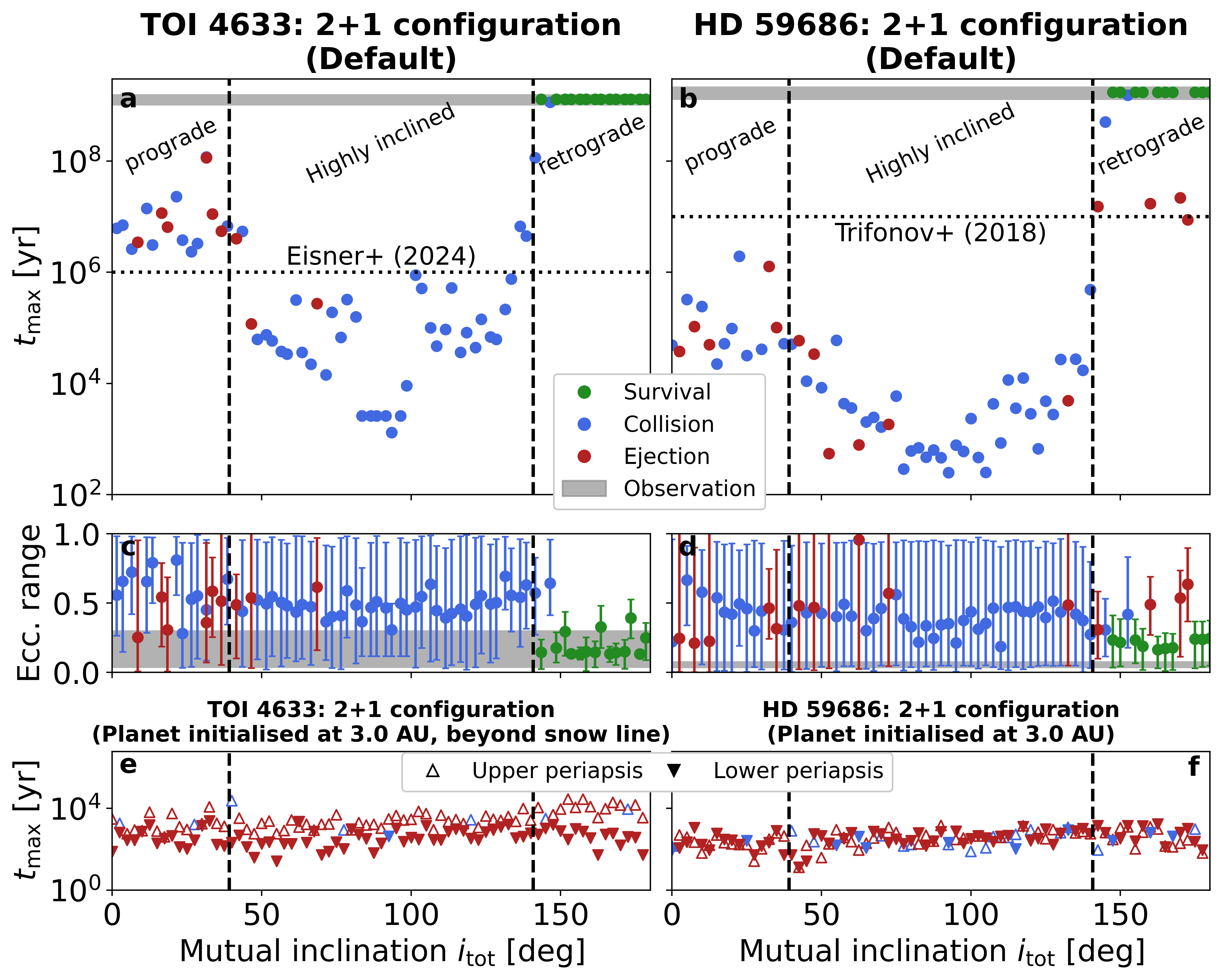}
    \caption{Evolutionary outcomes of the simulations as a function of the mutual inclination $i_\tot$ between the planetary and stellar orbits. Green markers indicate successful integrations until the maximum integration time $t_{\max}=t_\age$, blue markers indicate planetary collisions with either one of the stars prior to $t_\age$, and maroon markers indicate ejections of the planet. The left column shows TOI\,4633 in our default 2+1 configuration where planet b is ignored. The right column shows HD\,59686. The upper row (panels~a and~b) displays the maximum integration time until a certain evolutionary outcome is met. The middle row (panels~c and~d) indicates the range of orbital eccentricity (minimum, mean, and maximum) the planet attains throughout its evolution. In the lower row (panels~e and~f) the planet is initialised on a circular orbit at a semi-major axis of $3.0\,\au$ (corresponding to the snow line for TOI\,4633 \cite{hayashi1981,Ida2005}) and, within their one-sigma uncertainty, the binary semi-major axis $a_\B$ and eccentricity $e_\B$ are chosen to give the maximum (unfilled triangles) and minimum periapsis $r_{p,\B}=a_\B(1-e_\B)$ (filled triangles). Vertical dashed lines indicate the Kozai angles $i_\tot=39.2^\circ$ and $140.77^\circ$. Horizontal dotted lines show the maximum integration time assumed by \citet{Eisner2024} and \citet{Trifonov2018} to simulate TOI\,4633 and HD\,59686, respectively. Grey-shaded areas indicate the observational constraints on the age and the present-day orbital eccentricity of TOI\,4633 and HD\,59686, respectively.}
    \label{fig:TOI-4633}
\end{figure}

Figure~\ref{fig:TOI-4633}, panel~a, shows all simulation outcomes (survival in green, collision in blue, and ejection in red) as a function of the mutual inclination $i_\tot$ between the orbit of the planet and the stellar binary. The stability of a three-body system is dependent on the mutual inclination between the inner and outer system, where retrograde orbits are known to be more stable than prograde ones. This was first shown by \cite{Har72b}, and arises from Coriolis effect \cite{Ina80}, as we discuss below. Another dependence arises from secular von Zeipel-Lidov-Kozai \cite[vZLK;][]{Zeipel1910,Kozai1962,Lidov1962} effect, which can excite the eccentricities of the inner binary, making the system less stable, in particular between the so-called Kozai angles $40^\circ\lesssim i_{\tot} \lesssim 140^\circ$ \cite{per+12,Gri+17}. Due to these effects, the simulations exhibit three characteristic regimes. Between the Kozai angles the stellar companion exerts a torque causing large-amplitude eccentricity oscillations of the planetary orbit through vZLK effect, which most often leads to a collision with the host star or ejection from the system. The vZLK effect is suppressed for less inclined orbits at $i_\tot\lesssim40^\circ$ (prograde) and $i_\tot\gtrsim140^\circ$ (retrograde), in which case the planet survives for longer times. However, a striking asymmetry unravels for longer integration times where the large majority of retrograde simulations survives throughout the entire simulated age of TOI\,4633, whereas all prograde planets either collide with the star or get ejected beforehand. 
Notably for the case of prograde orbits, dynamical instabilities occur significantly later than $t_{\max}=1\,\rm Myr$. This effectively renders the standard stability assessment based on shorter integration times inapplicable to determine the stability of systems like TOI\,4633 \cite[e.g.,][]{Holman1999,Eisner2024}. 
A retrograde planetary orbit is further supported by its eccentricity evolution displayed in panel~c, which includes its minimum, maximum, and mean eccentricity throughout the integration. Only in the retrograde case do we find it to be largely consistent with its observed value $e_\Ac=0.117^{+0.186}_{-0.085}$. The different stability of near-coplanar prograde and retrograde orbits in hierarchical three-body systems is a well-known phenomenon, which is clearly visible in the case of TOI\,4633. This is because, in the frame rotating with the motion of the stellar binary, the planet experiences a Coriolis force \cite{Ina80}. The force's direction depends on the relative sense of rotation between binary orbit and planet orbit. Thus, it pushes the planet further away from its host in the co-rotating, prograde case, whereas pushing it closer in the counter-rotating, retrograde case, i.e., making the planet orbit around the host less or more stable, respectively \cite{Ina80}.

TOI\,4633 contains a second planet (TOI\,4633b) that was discovered through radial velocity variations to move around either of the stars at a much tighter orbit than TOI\,4633c \cite{Eisner2024}. We have tested various configurations including TOI\,4633b whose orbital properties are poorly constrained from observations, which we show in Supplementary Figure~\ref{fig:Supplementary_Figure1}. First, we arrange TOI\,4633 as a 3+1 quadruple with planet b orbiting star A, where we closely align the orbits of both planets but ensure that planet b is not seen as a transit (panels~a and~c), align it with the stellar binary TOI\,4633AB (panels~b and~d), and randomly orient planet b's orbit (panels~e and~g). Second, we consider a 2+2 quadruple with planet b orbiting star B, where we align its orbit with that of the stellar binary TOI\,4633AB (panels~f and~h) and draw a random orientation (panels~i and~k). In all cases, we keep varying the orbital orientation of planet c (see Methods). In these cases, we generally observe a larger scatter in the evolutionary outcomes so that some retrograde configurations do not survive, mostly because of planet b becoming unstable. Nevertheless, all studied cases result in similar retrograde-only stable configurations and we conclude that regardless of the orbital configuration of planet b, the quadruple TOI\,4633 could only survive until $t_{\rm max}=t_{\rm age}$ if planet c's orbit is retrograde with respect to the stellar binary. 

These findings relax if we assume the largest periapsis of the stellar binary allowed by the observational one-sigma uncertainties on its eccentricity $e_\B=0.91\pm0.03$ and semi-major axis $a_\B=48.6^{+4.4}_{-3.5}\,\au$. In Supplementary Figure~\ref{fig:Supplementary_Figure1}, panels~j and~l, we show our default 2+1 configuration initialised with $e_\B=0.88$ and $a_\B=53.0\,\au$, i.e., $r_{p,\B}=a_\B(1-e_\B)\approx7.5a_\Ac$. In this case, also all prograde planets ($0^\circ\lesssim i_{\rm tot}\lesssim40^\circ$) and even a few highly inclined systems within the Kozai angles ($40^\circ\lesssim i_{\rm tot}\lesssim140^\circ$) would survive. In order to further test the stability of prograde orbits within the observational uncertainty, we consider the prograde and co-planar configuration ($i_\tot=0.0^\circ$) and systematically vary the binary eccentricity and semi-major axis within the one-sigma uncertainty. As shown in Supplementary Figure~\ref{fig:Supplementary_Figure2} the prograde planet orbit would still get destabilised for most of the allowed parameter space and survives only if the binary periapsis is $r_{p,\B}\gtrsim5.9a_\Ac\approx5.0\,\au$. Following the Bayesian parameter estimation of \citet{Eisner2024} to construct posterior distributions for $a_\B$ and $e_\B$ from the observations of TOI\,4633 (see contours in Supplementary Figure~\ref{fig:Supplementary_Figure2}), we find an $80.0\,\%$ probability that $r_{p,\B}$ is below the threshold of $5.0\,\au$, where only retrograde planetary orbits are found to be stable. Thus, it is likely that the binary periapsis is indeed small enough to prevent any stable prograde configurations of the planet, though a wider companion allowing other configurations cannot be ruled out.
Regardless, either scenario~--~a sufficiently close companion ($r_{p,\B}\lesssim5.0\,\au$) which enforces a retrograde planetary orbit, or a wider companion ($r_{p,\B}\gtrsim5.0\,\au$) that also allows prograde planetary orbits~--~challenges our understanding of planet formation, as explained in the following.

We conducted additional simulations of TOI\,4633 in the 2+1 configuration where we initialised the planet on a circular orbit beyond the snow line at $a_\Ac=3.0\,\au$ \cite{hayashi1981,Ida2005}. Figure~\ref{fig:TOI-4633}, panel~e, shows that the planet rapidly destabilises mostly through ejections in less than about $10^4\,\rm yr$. These findings are nearly independent of the mutual inclination and even hold if we assume the observational upper bound on the binary periapsis (unfilled triangle markers). This timescale is much shorter than the planet could radially migrate inwards to its current orbit \cite{Tanaka2002}. Therefore, the planet could not have formed beyond the snow line in a circumprimary protostellar disc alongside an existing binary companion but would need to form much closer to its host, perhaps near its current position. Such in-situ formation at sub-snow-line distances is not expected in the standard core-accretion model \cite{Mizuno1980,Pollack1984,Lissauer1993}. Moreover, close binary companions with projected separations $\rho < 50\,\au$ significantly reduce both the occurrence rate and longevity of protoplanetary discs \cite{kraus12} and diminish exoplanet occurrence rates \cite{kraus16}. Additionally, a large binary eccentricity would further shorten the disc’s lifetime and spatial extent, making planet formation in such systems feasible only with a massive, low-eccentricity, self-gravitating disc, which is improbable \cite{xie10, raf15}. Alternatively, the stellar companion might have been captured as a result of stellar scattering in a dense environment, after the planet has formed. However, recent star-formation simulations \cite{Gen+25} show that although $\sim20\%$ of stars experienced an exchange interaction already during the star-formation, these mostly originated from disintegration of higher multiplicity systems, and not from single-formed stars which became binaries. Nevertheless, later exchanges in the cluster environment could occur on longer timescales.   This seems plausible for three reasons: i) In the absence of a perturbing stellar companion, the planet could have formed beyond the snow line and subsequently migrated to its current position. ii) The large eccentricity is a probable result of dynamical assembly of the binary \cite{Heggie1975,Ostriker1994,Mardling2001}. iii) If the planet orbit is indeed retrograde relative to the binary it would be a possible outcome of a star captured from a random direction \cite{Ostriker1994}. Conversely, if the star and planet had formed from the same gas clump, their orbital angular momenta would align, making a prograde orientation much more plausible.  
Other proposed mechanisms to form retrograde planets from long-term interactions with another more distant planet \cite{Naoz2011} or from planet-planet scatterings among p-type (circumbinary) planets \cite{Gong2018} seem implausible to explain TOI\,4633c's orbit. Another outer planet would have even less likely survived the presence of the binary companion, while Ref.~\cite{Gong2018} showed that the formation of retrograde s-type planets from p-type planet scatterings is suppressed for large binary eccentricities.

From the confirmed sample of discovered exoplanets, only HD\,59686 \citep{Ortiz2016} shares similar properties of a highly eccentric stellar binary ($a_\B\approx13.6\,\rm AU$, $e_\B\approx0.73$) and dynamical fragility ($a_\Ab\approx1.1\,\rm AU$, cf. Figure \ref{fig:s-type-Sample}). \citet{Trifonov2018}  conducted numerical integrations of HD\,59686 for up to $t_{\max}=10\,\rm Myr$ (not considering the possibility for planet-star collisions), finding it stable for retrograde orbits with $145^\circ\lesssim i_\tot\lesssim180^\circ$ and in a narrow region of prograde solutions locked in a secular apsidal alignment. However, unlike in TOI\,4633, the planet and stellar companion in HD\,59686 were discovered through radial-velocity measurements which leave the orbital orientation of the binary and planet less constrained (Supplementary Table~\ref{tab:orbital-parameters}) greatly increasing the allowed parameter space and making its stability analysis less complete. We conducted our analysis for HD\,59686 up to its estimated age $t_{\age}=1.73\pm0.47\,\rm Gyr$ \cite{Ortiz2016} (see Methods), and find qualitatively similar results as in TOI\,4633. The right column of Figure~\ref{fig:TOI-4633} shows that we only find retrograde solutions to be stable (albeit there is less agreement with the observed eccentricity range in panel~d), that the planet would rapidly disperse if it was placed further outside towards HD\,59686's snow line ($\sim9.7\,\rm\au$ \cite{Ortiz2016}), and Supplementary Figure~\ref{fig:Supplementary_Figure3} shows no stable prograde solution within the observational uncertainty of the semi-major axis and orbital eccentricity of the stellar binary. While the same formation scenarios like above may also explain the origin of the planet in the hostile binary HD\,59686AB, \citet{Ortiz2016} highlight another plausible alternative. Unlike in TOI\,4633, the binary is composed of evolved stars, with the companion likely being a white dwarf. Its progenitor star must have initiated a mass-transfer episode onto HD\,59686A. This would have led to an accretion disk around it from which the planet may have formed as a second-generation circumstellar planet \cite{Perets2010,Tutukov2012}, alleviating the need for long-term stability over $t_\age$. While the details of this scenario are uncertain, we summarise that TOI\,4633 harbours the only known planet whose origin could be either explained by sub-snow-line in-situ formation or late dynamical assembly of the binary companion. 

Our analysis of TOI\,4633 could serve as a blueprint to study the orbits of many transiting s-type planets which are soon detectable with the spacecraft \textit{Plato} and whose hosts could also reveal a fully astrometrically resolvable companion in the coming data release of \textit{Gaia} (DR4). After its launch in late 2026, \textit{Plato} is designed to detect about $\sim4600$ planets transiting stellar hosts with \textit{V}-band magnitudes $V\leq13$ at periods $T\lesssim10^3\,\rm days$ \cite{PLATO2017}. We find that in the latest \textit{Gaia} data release (DR3) a fraction of $4.0$ percent of all bright stars ($G\leq13$) in the field-of-view of the first \textit{Plato} Long-duration Observation Phase (LOPS2) have stellar companions on astrometrically resolved orbits with $T_\B\lesssim10^3\,\rm days$. This fraction might increase in \textit{Gaia} DR4 (after mid-2026), potentially up to the close binary fraction of $10$~--~$20$ percent observed for low-mass stars \cite{Offner}. Unless a close stellar binary companion strongly suppresses the presence of planets, we can therefore plausibly expect tens to a few hundred discoveries of s-type planets in close binary stars, where, like in TOI\,4633, \textit{Plato} transits and \textit{Gaia} astrometry only leave $i_\tot$ fundamentally unconstrained. This presents an opportunity to test the retrograde orbit hypothesis via radial velocity observations and the Rossiter-McLaughlin effect \cite{Rossiter1924,McLaughlin1924,Albrecht2022} for potentially hundreds of systems discovered by \textit{Plato}. As demonstrated in this work, our method could serve as a blueprint to uncover their orbital state and to constrain their evolutionary past. 

\section*{Methods}
We use the direct $N$-body code {\tt Rebound} \cite{rebound} to simulate different orbital configurations of TOI\,4633 using the built-in Gragg-Bulirsch-Stoer integrator \citep{Gragg1965,Bulirsch1966} with relative and absolute tolerances of $10^{-12}$. We have also tested evolving several systems (including the long-integration survivors) with the more accurate (albeit computationally more expensive) Integrator with Adaptive Step-size control at 15th order (IAS15) \cite{Everhart1985,Rein2015} without finding a difference in the evolutionary outcome. In order to initialise our default 2+1 configuration (Figure~\ref{fig:TOI-4633}, panel~a and~c), we adopt all observed mean values reported in Supplementary Table~\ref{tab:orbital-parameters} for the masses $m_\A$, $m_\B$, and $m_{\rm c}$, semi-major axes $a_\B$ and $a_\Ac$, orbital eccentricities $e_\B$ and $e_\Ac$, as well as orbital angles in the observer frame $\omega_\B$ (binary's argument of pericentre), $\Omega_\B$ (binary's longitude of ascending node), $i_\B$ (binary's inclination relative to the line of sight), and $i_\Ac$ (planet's inclination relative to the line of sight). In each simulation we set the observationally unconstrained longitude of the ascending node of the planet to $\Omega_\Ac=0^\circ,5^\circ,\dots,355^\circ$ (summing up to 72 simulations in total), resulting in a different mutual inclination $i_{\rm tot}$ as
\begin{equation}
    \cos i_\tot=\cos i_\B\cos i_\Ac + \sin i_\B\sin i_\Ac\cos(\Omega_\B-\Omega_\Ac).\label{eq:itot}
\end{equation}
Since the large measurement uncertainties of the planet's argument of pericentre $\omega_\Ac$ make it effectively unconstrained too, we opt to sample it randomly between $0$ and $360^\circ$ in each simulation. Furthermore, we pick random initial phases in time by sampling the mean anomalies of the stellar and planetary orbits from a uniform distribution between $0$ and $360^\circ$. Each simulation is terminated after it reaches a maximum integration time $t_{\max}=1.3\,\rm Gyr$, if the distance between TOI\,4633A and TOI\,4633c exceeds $100\,\au$, or if the distance between TOI\,4633c to either of the stars gets smaller than their combined radii $r_\A+r_{\rm c}$ and $r_\B+r_{\rm c}$, respectively, which are adopted from the observations (Supplementary Table~\ref{tab:orbital-parameters}). In order to justify our limit of $100\,\au$, we repeat the simulation of several ejected systems and verified that they indeed reach a much larger distance of $10^3\,\au$.

Furthermore, we conduct simulations in which we take the presence of planet TOI\,4633b into account. For those, we initialise 72 simulations exactly like above and add planet b as follows. In the first configuration, we put it on an orbit around the host star A with $i_\Ab=88.0^\circ$ and $\Omega_\Ab=\Omega_\Ac$ (Figure~\ref{fig:Supplementary_Figure1}, panels~a and~c). This 3+1 configuration is adopted from \citet{Eisner2024} and represents a near-alignment with the orbit of c where the slight difference in inclination is to ensure that planet b is not observed as a transit. Second, we align its orbit with the stellar binary, i.e., $i_\Ab=i_\B$ and $\Omega_\Ab=\Omega_\B$ (panels~b and~d). Third, we keep this 3+1 configuration but randomise the spatial orientation of the orbit of planet b, i.e., we draw $\Omega_\Ab$, $\omega_\Ab$, and $\cos i_\Ab$ from uniform distributions (panel~e and~g).  The set-up of the latter two configurations is repeated but with planet b orbiting the other star TOI\,4633B (panels~f and~h and panels~e and~g, respectively). In each of the configurations we randomise the mean anomaly and $\omega_\Ab$ of the planet, calculate its mass as $m_{\rm b}\times\sin i_\Ab=106.8\,\rm M_\oplus$, and compute its semi-major axis from Kepler's Third law
\begin{equation}\label{eq:Kepler}
    a_\Ab={{\sqrt[{3}]{\frac {Gm_\tot T_\Ab^{2}}{4\pi ^{2}}}}},
\end{equation}
where $G$ is the gravitational constant and the total mass is $m_\tot=m_\A+m_{\rm b}$ for the 3+1 configurations and $m_\tot=m_\B+m_{\rm b}$ for the 2+2 configurations. In all of these quadruple simulations, we also terminate the integration if planet b gets ejected or collides with one of the stars. Lacking a radius estimate for the planet we assume $r_{\rm b}=10\,\rm R_\oplus$ which is typical for planets of that mass \cite{Muller2024}.

Moreover, we explored auxiliary 2+1 configurations of TOI\,4633 with different values for the binary semi-major axis and eccentricity within the observational one-sigma uncertainty $a_\B=48.6^{+4.4}_{-3.5}\,\au$ and $e_\B=0.91\pm0.03$, respectively. Firstly, we have initialised the largest binary periapsis, i.e., $a_\B=53.0\,\au$ and $e_\B=0.88$ and choose the other parameters as described above (panels~j and~l). Secondly, we fix the relative mutual inclination $i_\tot$ to zero and explore the entire uncertain parameter space by a grid of simulations with step sizes $\Delta a_\B=0.4\,\au$ and $\Delta e_\B=0.01$ (Supplementary Figure~\ref{fig:Supplementary_Figure2}).

Lastly, we investigated the 2+1 configuration of HD\,59686 composed of the host star A, companion star B, and planet b (Figure~\ref{fig:TOI-4633}, right column and Supplementary Figure~\ref{fig:Supplementary_Figure3}). Since the orbital orientation of the stellar binary and planet are less constrained, we opt to choose a reference frame in which $\Omega_\B=\Omega_\Ab=i_\Ab=0^\circ$ and set in each simulation the binary inclination to $i_\B=i_\tot=0.0^\circ,2.5^\circ,\dots,180.0^\circ$. Component masses, eccentricities, binary semi-major axis, and planet orbital period are adopted from the observational mean values \cite[][Supplementary Table~\ref{tab:orbital-parameters}]{Ortiz2016}. Assuming that we observe the system from a isotropically random direction, we divide the companion mass and planet mass by the sines of the line-of-sight inclinations, respectively, and calculate the planet semi-major axis from Eq.~\eqref{eq:Kepler}. The remaining angles are randomly sampled.

\renewcommand{\tablename}{Supplementary Table}
\begin{table}[]
    \centering
    \begin{tabular}{lcc}
    \toprule
        Parameters & TOI\,4633 \citep{Eisner2024} & HD\,59686 \citep{Ortiz2016}  \\
        \midrule
       \textit{Outer orbit of the stars:} & & \\
       Host mass $m_\A$ $[\msun]$ & $1.10\pm 0.06$ & $1.9\pm0.2$ \\
       Companion mass $m_\B$ $[\msun]$ & $1.05\pm 0.06$ & $0.5293^{+0.0011}_{-0.0009}\times(\sin i_{\B})^{-1}$  \\
       Age $t_{\age}$ $[\rm Gyr]$ & $1.3\pm0.3$ & $1.73\pm0.47$ \\
       Host radius $r_\A$ $[\rm R_\odot]$ & $1.05\pm 0.05$ & $13.2\pm0.3$ \\
       Semi-major axis $a_\B$ $[\au]$ & $48.6^{+4.4}_{-3.5}$ & $13.56^{+0.18}_{-0.14}$ \\
       Orbital period $T_\B$ $[\rm yr]$ & $231^{+32}_{-24}$ & $31.98^{+0.64}_{-0.47}$ \\
       Eccentricity $e_\B$ & $0.91\pm0.03$ & $0.729^{+0.004}_{-0.003}$ \\
       Argument of pericentre $\omega_\B$ $[\rm deg]$ & $110.5\pm2.1$ & $149.4\pm0.2$ \\
       Longitude of ascending node $\Omega_\B$ $[\rm deg]$ & $123.5^{+3.3}_{-2.9}$ & n/a \\
       Inclination $i_\B$ $[\rm deg]$ & $90.1\pm0.4$ & n/a \\
       \midrule
       \textit{Orbit of the inner planet:} & TOI\,4633b\footnotemark[1] & HD\,59686b \\
       Planet minimum mass $m_{\rm b}\sin i_\Ab$ $[\rm M_\oplus]$ & $106.8^{+13.0}_{-12.8}$ & $2199.18^{+57.20}_{-76.27}$\\
       Orbital period $T_\Ab$ $[\rm days]$ & $34.15\pm0.15$ & $299.36^{+0.26}_{-0.31}$ \\
       Eccentricity $e_\Ab$ & $0.096^{+0.102}_{-0.065}$ & $0.05^{+0.03}_{-0.02}$ \\
       Argument of pericentre $\omega_\Ab$ $[\rm deg]$ & $-43.9^{+104.8}_{-72.8}$ & $121^{+28}_{-24}$  \\
       Longitude of ascending node $\Omega_\Ab$ $[\rm deg]$ & n/a & n/a \\
       Inclination $i_\Ab$ $[\rm deg]$ & $\not\approx90.0$ (not transiting) & n/a \\
       \midrule
       \textit{Orbit of the outer planet:} & TOI\,4633c & \\
       Planet mass $m_{\rm c}$ $[\rm M_\oplus]$ & $47.8^{+27.6}_{-23.8}$ & \\
       Planet radius $r_{\rm c}$ $[\rm R_\oplus]$ & $3.2^{+0.20}_{-0.19}$  \\
       Semi-major axis $a_\Ac$ $[\au]$ & $0.847\pm0.061$ &  \\
       Critical semi-major axis\footnotemark[2] $a_{\rm crit}$ $[\au]$ & $0.47^{+0.35}_{-0.28}$ &  \\
       Orbital period $T_\Ac$ $[\rm days]$ & $271.9445^{+0.0039}_{-0.0040}$ &  \\
       Eccentricity $e_\Ac$ & $0.117^{+0.186}_{-0.085}$ & \\
       Argument of pericentre $\omega_\Ac$ $[\rm deg]$ & $-21^{+131}_{-108}$ &   \\
       Longitude of asc. node $\Omega_\Ac$ $[\rm deg]$ & n/a & \\
       Inclination $i_\Ac$ $[\rm deg]$ & $89.888^{+0.069}_{-0.064}$ & \\
       \botrule
    \end{tabular}
    \footnotetext[1]{Observations do not tell whether the planet TOI\,4633b is orbiting star TOI\,4633A or B \cite{Eisner2024}. Nonetheless, we denote its parameters with subscripts ``b" and ``Ab" for the ease of reading.}
    \footnotetext[2]{$a_{\rm crit} = \left[ (0.464 \pm 0.006) + (-0.380 \pm 0.010) \mu + (-0.631 \pm 0.034)e_\B + (0.586 \pm 0.061) \mu e \right. \\
+ (0.150 \pm 0.041)e_\B^2 + (-0.198 \pm 0.074) \mu e_\B^2 \left. \right] a_\B
$ where $\mu=m_\B/(m_\A+m_\B)$ \cite{Holman1999}.}
    \caption{Orbital parameter values of the two planetary systems TOI\,4633 and HD\,59686 inferred by \citet{Eisner2024} and \citet{Ortiz2016}, respectively. TOI\,4633 harbours two known planets, HD\,59686 just one.}
    \label{tab:orbital-parameters}
\end{table}

\renewcommand{\figurename}{Supplementary Figure}
\setcounter{figure}{0}
\begin{figure}
    \centering
    \includegraphics[width=\linewidth]{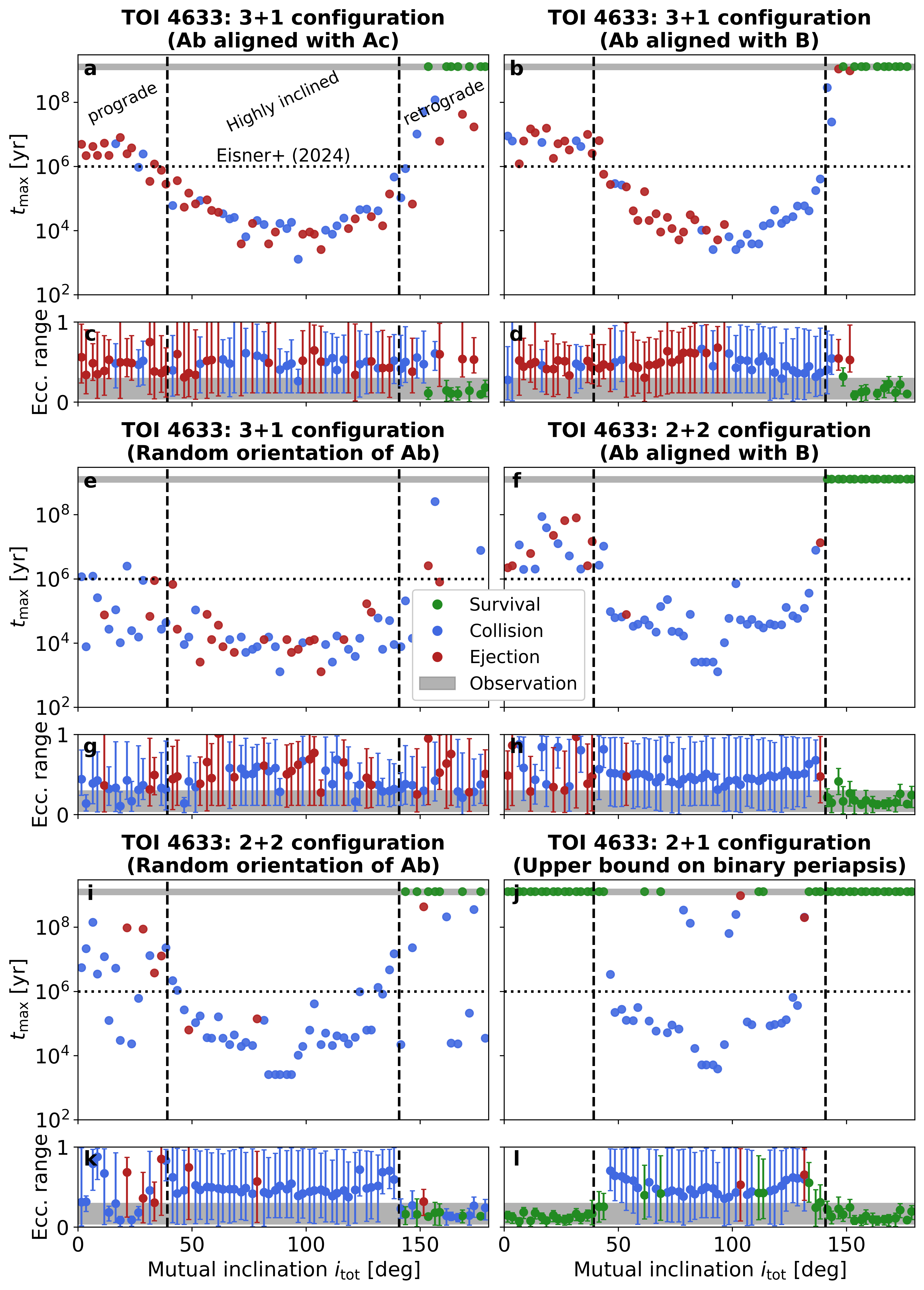}
    \caption{Same as Figure~\ref{fig:TOI-4633} for different configurations of TOI\,4633 (see description in the Methods).}
    \label{fig:Supplementary_Figure1}
\end{figure}

\begin{figure}
    \centering
    \includegraphics[width=\linewidth]{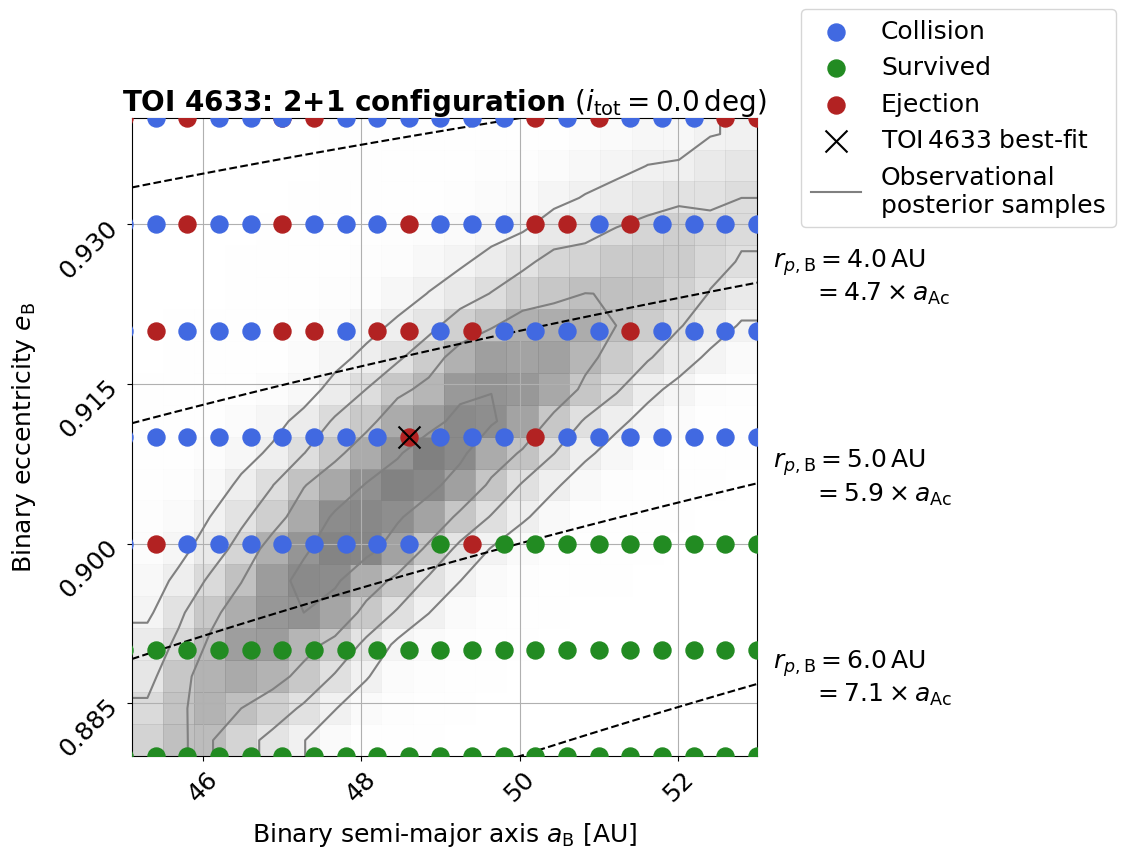}
    \caption{Evolutionary outcomes within the observational one-sigma uncertainty of the binary semi-major axis $a_\B=48.6^{+4.4}_{-3.5}\,\au$ and orbital eccentricity $e_\B=0.91\pm0.03$. We consider our default 2+1 configuration and assume TOI\,4633c to be on a prograde orbit ($i_\tot=0.0^\circ$). As in Figure~\ref{fig:TOI-4633}, green, blue, and red markers indicate the survival of the system, the collision of the planet with a star, and a planetary ejection. The black cross indicates the best observational fit to the orbital binary parameters of TOI\,4633AB at $a_\B=48.6\,\au$ and $e_\B=0.91$, while grey (0.5, 1, 1.5, 2)-sigma equivalent contours indicate the two-dimensional posterior distribution of $a_\B$ and $e_\B$ from the observational data of TOI\,4633 \citep[][see Fig.~7]{Eisner2024}. The boundaries of the plot match the observational one-sigma uncertainties of the marginal posterior distributions of $a_\B$ and $e_\B$. Dashed contour lines show constant values for the binary periapsis at $r_{p,\B}=3.0$, $4.0$, $5.0$, and $6.0\,\au$.}
    \label{fig:Supplementary_Figure2}
\end{figure}

\begin{figure}
    \centering
    \includegraphics[width=\linewidth]{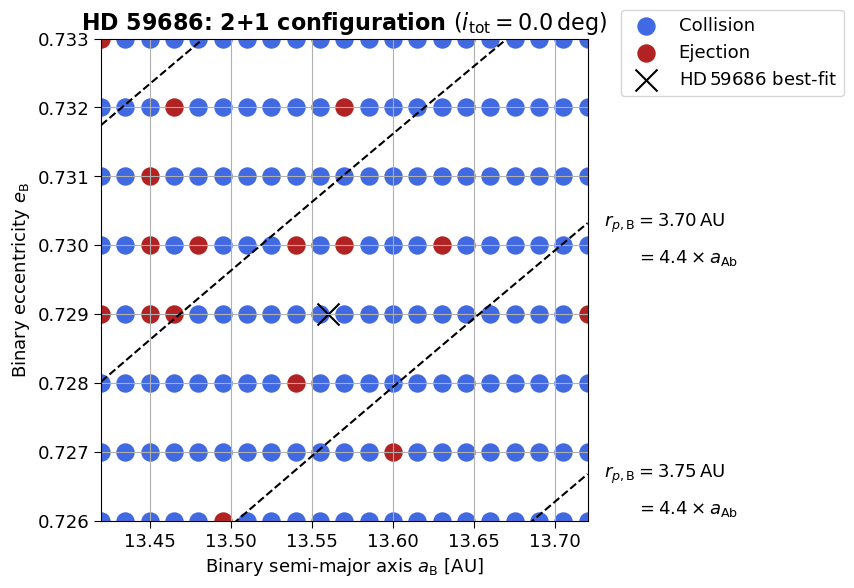}
    \caption{Same as Supplementary Figure~\ref{fig:Supplementary_Figure2} for HD\,59686.}
    \label{fig:Supplementary_Figure3}
\end{figure}


\subsection*{Sources of precession and its effects on the dynamics}
In realistic systems, the planet TOI\,4633c might be stabilised by relativistic and tidal effects which cause its orbit to precess quenching the perturbation from the stellar companion \cite[e.g.,][]{Liu2015}. These effects are not included in our simulations. However, we show below that they are expected to be less important than the precession in the vZLK dynamics and the precession arising from the presence of the inner planet TOI\,4633b (in the 3+1 configuration) highlighting the robustness of our results. 

\subsubsection*{Schwarzschild precession from general relativity}
The relativistic Schwarzschild precession can be compared to the secular precession rate from the outer stellar companion. The dimensionless precession rate is given by
$\epsilon_{\rm GR}$ \citep{lml15,man22}:
\begin{equation}
 \epsilon_{{\rm GR}}  \equiv \left|\frac{\dot{\omega}_{{\rm Ac,GR}}}{\dot{\omega}_{{\rm Ac,vZLK}}}\right|= \frac{3m_{{\rm bin}}}{m_{{\rm out}}}\frac{a_{\rm B}^{3}(1-e_{\rm B})^{3/2}}{a_{\rm Ac}^{3}}\frac{Gm_{{\rm bin}}}{a_{\rm Ac}c^{2}},
\end{equation}
where $m_{\rm bin} = m_A + m_{\rm c}$, $m_{\rm out}=m_{\rm B}$. For $\epsilon_{\rm GR} \gg 1$ the effects from the stellar companion are negligible, while for $\epsilon_{\rm GR} \ll 1$ the relativistic effect is small. Nevertheless, for small $\epsilon_{\rm GR} \ll 1$ the maximal eccentricity is limited by
$e_{{\max}}=\sqrt{1-8\epsilon_{{\rm GR}}^{2}/9}$, even for optimal mutual inclination. 
For the parameters of planet c we have $\epsilon_{{\rm GR}}=5\times10^{-4}$ and $1-e_{\max}\approx 10^{-6}$ which
is insignificant. We conclude that the extra precession from general relativity does not affect the long-term dynamics of the TOI\,4633 system.

\subsubsection*{Precession from tidal bulges}
Tidal interactions can raise equilibrium bulges on the planets and stars. This deviation from spherical symmetry will also lead to precession. The relative strength of the precession is parametrised by the apsidal motion constants $k_{\rm 1 \star}$ and $k_{\rm 1 p}$ for the stars and planets respectively. Similarly, we can define the relative precession rate compared to the vZLK one 
\cite{lml15, man22}
\begin{equation}
\epsilon_{{\rm tide}} \equiv \left|\frac{\dot{\omega}_{{\rm Ac,tide}}}{\dot{\omega}_{{\rm Ac,vZLK}}}\right| =\frac{15m_{{\rm bin}}a_{\rm B}^{3}(1-e_{\rm B})^{3/2}(k_{\rm 1 \star}m_{\rm A}^{2}R_{\rm A}^{5}+k_{\rm 1p}m_{\rm c}^{2}R_{\rm Ac}^{5})}{a_{\rm Ac}^{8}m_{\rm A}m_{\rm c}m_{\rm B}}
\end{equation}
where the two terms measure the tidal bulges raised on the star A
and the planet c. The apsidal motion constants are $k_{\rm 1 \star}=0.014,\,k_{\rm 1 p}=0.1$
(e.g. \cite{gp22}) and the radii are $r_\A=1.05\,\rm R_{\odot}$, $r_{\rm c}=0.05\,\rm R_{\odot}$. For the parameters of planet c, we have $\epsilon_{{\rm tide}}\approx2\times10^{-4}$. The maximal eccentricity can be solved numerically, yielding $e_{{\max}}\approx0.92$. 
Similarly, for planet b, tides are more important, yielding $\epsilon_{{\rm tide}}\approx4.16$ and $e_{{\max}}\approx0.43$, which could contribute to the stability on the inner planet.
We conclude that tidal bulges may have greater importance than the Schwarzschild effect, but still could be safely ignored for the duration of the dynamical evolution of the outer planet. 

\subsubsection*{Precession from inner mass quadrupole}

The inner planet may also affect the orbit of planet c. In order to estimate this effect, we first define the ``effective Laplace radius" \cite{grilai18} which measures the relative importance of the inner quadrupole moment relative to the outer quadrupole perturbation from the binary companion:
\begin{equation}
 r_{{\rm L}}=\left(\frac{m_{{\rm A}}m_{{\rm b}}}{2m_{{\rm B}}(m_{{\rm A}}+m_{{\rm b}})}a_{{\rm Ab}}^{2}a_{{\rm B}}^{3}(1-e_{{\rm B}}^{2})^{3/2}\right)^{1/5}.
\end{equation}
 For our case, we have $r_{\rm L} \approx 0.615\,\rm\au$, which is slightly smaller than the planet c's current location. The relative precession rate of planet c compared to the vZKL rate is $\epsilon_{\rm rot} = 1.5 (a_{\rm Ac}/r_{\rm L})^{-5}\approx 0.3$. The maximal eccentricity can also be computed via  (e.g. \cite{gri20}) $e_{\max} = 1 - 2 \epsilon_{\rm rot}^{2/3}/9 \approx 0.9$.

We conclude that the inner mass quadrupole from the inner planet b is the most important effect, but still, the maximal eccentricity can reach very high values of almost $0.9$. We simulate four-body systems that capture this effect in Supplementary Figure~\ref{fig:Supplementary_Figure1} and find similar results when only retrograde planetary orbits are stable for $10^9\,\rm yr$. Moreover, since lower eccentricity is required for a close encounter with the inner planet, more systems become unstable on average and the instability times are somewhat shorter for the prograde orbits.

\subsubsection*{Precession of the outer orbit due to planet c}

Similarly, we can estimate the precession timescale for the binary stellar orbit as seen in Figure~\ref{fig:examples}. The precession timescale is a product of the mass ratios between the planets and the stars and the square of the semi-latus rectum of the outer orbit to the planet's semi-major axis:

\begin{equation}   
T_{\rm prec, B} \approx T_{\B}\times \frac{m_{\rm A} + m_{\B}}{m_{\Ac}} \frac{a_{\B} (1-e_{\B}^2)}{a_{\Ac}^2} \approx 170\,\rm Myr.
\end{equation}
The timescale is consistent with the observed precession of a few degrees during the integration time reported in Figure~\ref{fig:examples}.

\section*{Acknowledgements}
This research has made use of data obtained from or tools provided by the portal \url{exoplanet.eu} of The Extrasolar Planets Encyclopaedia \cite{Encyclopaedia}. J.S., S.J., and S.E.d.M.~acknowledge funding from the Netherlands Organisation for Scientific Research (NWO), as part of the Vidi research program BinWaves (project No. 639.042.728, PI: de Mink). C.J.~acknowledges funding from the Royal Society through the Newton International Fellowship funding scheme (project No. NIF$\backslash$R1$\backslash$242552). Computations were performed on the HPC system Raven at the Max Planck Computing and Data Facility.

\section*{Author contributions}
J.S.~conceived the study, performed the numerical and analytical computations, and wrote the manuscript. E.G. contributed to the numerical computations and analytical calculations. C.J.~and N.L.E.~provided observational expertise of TOI\,4633. H.B.P. contributed to the stability analysis and relation of stability and post-snow-line planet formation. All authors contributed to the interpretation of the results and reviewed and edited the manuscript.

\section*{Declarations}
The authors declare no competing interests.


\bibliography{sn-bibliography}

\end{document}